\documentstyle[fleqn,11pt]{article}
\begin{document}
\bigskip
\parindent 1.4cm
\large

\begin{center}

{\Large \bf Bohmian Trajectories of Airy Packets
}
\end{center}
\vspace{1.0cm}
\begin{center}
{Antonio B. Nassar${^1}$ and Salvador Miret-Art\'es${^2}$}
\end{center}
\begin{center}
{\it ${^1}$Science Department, Harvard-Westlake School,
\par 3700 Coldwater Canyon, Studio City, 91604, USA
\par ${^1}$Department of Sciences, University of California,
Los Angeles, Extension Program
\par 10995 Le Conte Avenue, Los Angeles, CA 90024, USA
\par ${^2}$ Instituto de F\'isica Fundamental, Consejo Superior de
Investigaciones Cient\'ificas, Serrano 123, 28006 Madrid, Spain
}
\end{center}
\vspace{1.0cm}
\par
\begin{center}
{\bf Abstract}
\end{center}

The discovery of Berry and Balazs in 1979 that the free-particle Schr\"odinger equation allows a non-dispersive and accelerating Airy-packet solution has taken the folklore of quantum mechanics by surprise. Over the years, this intriguing class of wave packets has sparked enormous theoretical and experimental activities in related areas of optics and atom physics. Within the Bohmian mechanics framework, we present new features of Airy wave packet solutions to Schr\"odinger equation with time-dependent quadratic potentials. In particular, we provide some insights to the problem by calculating the corresponding Bohmian trajectories. It is shown that by using general space-time transformations, these trajectories can display a unique variety of cases depending upon the initial position of the individual particle in the Airy wave packet. Further, we report here a myriad of nontrivial Bohmian trajectories associated to the Airy wave packet.
These new features are worth introducing to the subject's theoretical folklore in light of the fact that the evolution of a quantum mechanical Airy wave packet governed by the Schr\"odinger equation is analogous to the propagation of a finite energy Airy beam satisfying the paraxial equation. Numerous experimental configurations of optics and atom physics have shown that the dynamics of Airy beams depends significantly on initial parameters and configurations of the experimental set-up.

\vspace{1.0cm}
\vspace{0.1cm}
\par E-mail: nassar@ucla.edu
\vspace{1.0cm}

\pagebreak

The discovery of Berry and Balazs in 1979  \cite{berry} that the Schr\"odinger equation with linear potential allows a non-dispersive and accelerating Airy wave packet solution has taken the folklore of quantum mechanics by surprise. They have shown that for the Schr\"odinger equation

\begin{eqnarray} \label{eq:berry2}
i\hbar\frac{\partial \psi (x,t)}{\partial t}=-\frac{{{\hbar }^{2}}}{2m}\frac{{{\partial }^{2}}\psi (x,t)}{\partial {{x}^{2}}}-(F(t)x)\psi (x,t),
\end{eqnarray}
a unique solution is given by the Airy wave packet probability density:
\begin{eqnarray} \label{eq:berry1}
{{\left| \psi (x,t) \right|}^{2}}=A_{i}^{2}\left\{ \frac{B}{{{\hbar }^{2/3}}}\left[ x-\frac{1}{m}\int\limits_{0}^{t}{d\tau F(\tau )(t-\tau )}-\frac{{{B}^{3}}{{t}^{2}}}{4{{m}^{2}}} \right] \right\}.
\end{eqnarray}

Over the years, this intriguing class of wave packets has sparked a considerable resurgence on research on diffraction theory and experiments on a classical analogue in optics of nondiffracting beams.\cite{siviloglou}-\cite{besieris2} Despite its numerous applications in electrodynamics, optical theory, solid state physics, radiative transfer, semiconductors in electric fields, Airy functions \cite{vallee} have found to be a distinctive solution in quantum mechanics but only for linear potentials.\cite{besieris2,unni,green}

Within the Bohmian mechanics framework,\cite{bohm}-\cite{bernstein} we present in this work new features of Airy wave packet solutions to Schr\"odinger equation with time-dependent quadratic potentials. In particular, we provide some insights to the problem by calculating its Bohmian trajectories. It is shown that by using general space-time transformations, these trajectories can display a unique variety of cases depending upon the initial position of the individual particle in the Airy wave packet. These results can further common mathematical similarities between Sch\"rodinger equation and the paraxial equation of diffraction and pave the way toward the discovery of new experimental observations. In particular, we report here that for a force-free Airy packet a dispersive and non-accelerating trajectory is also possible. These new features are worth adding to the subject's theoretical folklore in light of the already numerous experimental configurations of optics and atom physics. 
These new features are worth introducing to the subject's theoretical folklore in light of the fact that the evolution of a quantum mechanical Airy wave packet governed by the Schr\"odinger equation is analogous to the propagation of a finite energy Airy beam satisfying the paraxial equation. Numerous experimental configurations of optics and atom physics have shown that the dynamics of Airy beams depends significantly on initial parameters and configurations of the experimental set-up.

Our problem is defined by two evolution equations: the time-dependent equation Schr\"odinger equation for the wave function $\psi(x,t)$
\begin{eqnarray} \label{eq:schroedinger1}
i\hbar \frac{\partial \psi (x,t)}{\partial t}=-\frac{{{\hbar }^{2}}}{2m}\frac{{{\partial }^{2}}\psi (x,t)}{\partial {{x}^{2}}}+\left( \frac{1}{2}m{{\omega }^{2}}(t){{x}^{2}}-F(t)x \right)\psi (x,t)
\end{eqnarray}
and the first-order guiding equation for $x(t)$:
\begin{eqnarray} \label{eq:ximag}
{{\dot{x}}_{i}}(t)=\frac{\hbar }{m}Im{{\left. \left( \frac{\partial }{\partial x}\log \psi (x,t) \right) \right|}_{x={{x}_{i}}(t)}},
\end{eqnarray}
which constitutes the simplest first-order evolution equation for the position of the particle that is compatible with the Galilean (and time-reversal) covariance of the Schrödinger evolution. Our main objective in this work is to solve Equation (\ref{eq:ximag}) for the Bohmian trajectories of an evolving {\it ith} particle of the Airy wave packet ensemble with an initial position ${x}_{oi}$.

A general solution to Equation (\ref{eq:schroedinger1}) can be found through a proper time rescaling of the space variables and introducing new times.\cite{nassar,ray,burgan} This new group of transformations reduce Equation (\ref{eq:schroedinger1}) to the case of the problem of a free-particle type motion. Although different techniques have been introduced previously, the scale and phase transformations presented here yield a simpler physical meaning to the mathematical protocol. To this end, we introduce the extended space-time transformations

\begin{eqnarray} \label{eq:wavetrans}
\psi (x,t)=\frac{1}{\sqrt{\delta (t)}}{{\exp }^{{}}}\left( \frac{i{{\phi}_{1}}(x,t)}{\hbar } \right){{\psi }_{1}}(x,t),
\end{eqnarray}

\begin{eqnarray} \label{eq:x}
x'=\frac{1}{\delta (t)}\left( x-X(t) \right),
\end{eqnarray}
and
\begin{eqnarray} \label{eq:t}
{t}'=\int\limits_{0}^{t}{\frac{d\tau }{{{\delta }^{2}}(\tau )}}.
\end{eqnarray}

These transformations represent a scale and phase transformation on the wave function and a scale transformation on space and time along with a space translation. In particular, Equation (\ref{eq:x}) is a Galilean-type transformation. We find after lengthy but straightforward calculations that
Equation (\ref{eq:schroedinger1}) reduces to

\begin{eqnarray} \label{eq:schroedinger2}
i\hbar\frac{\partial {{\psi }_{1}}({x}',{t}')}{\partial {t}'}=-\frac{{{\hbar }^{2}}}{2m}\frac{{{\partial }^{2}}{{\psi }_{1}}({x}',{t}')}{\partial x{{'}^{2}}}+{{f}_{1}}({t}'){{\psi }_{1}}({x}',{t}')
\end{eqnarray}
where
\begin{eqnarray} \label{Ft}
{{f}_1}({t'})=\frac{m}{2}[{{\dot{\delta }}^{2}}{{X}^{2}}-{{\delta }^{2}}{{\dot{X}}^{2}}-2X\dot{X}\delta \dot{\delta }]+m{{\delta }^{2}}{{f}_o}(t').
\end{eqnarray}

By performing the phase change
\begin{eqnarray} \label{phasechange}
{{\psi }_{1}}(x',t')={{\psi }_{2}}(x',t')\exp \left( -\frac{i}{\hbar }\int\limits_{0}^{t'}{{{f}_{1}}(t'')dt''} \right),
\end{eqnarray}
Equation (\ref{eq:schroedinger2}) reduces further to
\begin{eqnarray} \label{freeSch}
i\hbar\frac{\partial {{\psi }_{2}}({x}',{t}')}{\partial {t}'}=-\frac{{{\hbar }^{2}}}{2m}\frac{{{\partial }^{2}}{{\psi }_{2}}({x}',{t}')}{\partial x{{'}^{2}}}.
\end{eqnarray}

The final solution to Equation (\ref{eq:schroedinger1}) then reads

\begin{eqnarray} \label{eq:finalsol}
\psi (x,t)=\frac{1}{\sqrt{\delta (t)}}{{A}_{i}}\left\{ \frac{B}{{{\hbar }^{2/3}}}\left[ \frac{\left[ x-X(t) \right]}{\delta (t)}-\frac{{{B}^{3}}}{4{{m}^{2}}}{{\left( \int\limits_{0}^{t}{\frac{d\tau }{{{\delta }^{2}}(\tau )}} \right)}^{2}} \right] \right\}\exp \frac{i}{\hbar }\phi(x,t),
\end{eqnarray}
where\cite{phase}
\begin{eqnarray} \label{eq:Sfunc}
\frac{\partial \phi (x,t)}{\partial x}=\frac{\partial }{\partial x}\left[ {{\phi }_{1}}(x,t)+{{\phi }_{2}}(x,t) \right]=\frac{m\dot{\delta }(t)}{\delta (t)}[x-X(t)]+m\dot{X}(t)+\frac{{{B}^{3}}}{2m\delta(t)}\int\limits_{0}^{t}{\frac{d{t}'}{{{\delta }^{2}}({t}')}}
\end{eqnarray}
and the auxiliary functions $X(t)$ and $\delta(t)$ obey
\begin{eqnarray} \label{eq:xeq}
\ddot{X}(t)+{{\omega }^{2}}(t)X(t)=\frac{F(t)}{m}
\end{eqnarray}
and
\begin{eqnarray} \label{eq:deltaeq}
\ddot{\delta }(t)+{{\omega }^{2}}(t)\delta (t)=0.
\end{eqnarray}

Now, with the help of Equation(\ref{eq:Sfunc}), Equation (\ref{eq:ximag}) can be recast as
\begin{eqnarray} \label{xdot}
{{\dot{x}}_{i}}(t)=\frac{1}{m}{{\left. \left( \frac{\partial \phi (x,t)}{\partial x} \right) \right|}_{x={{x}_{i}}(t)}}=\dot{X}(t)+\left[ x-X(t) \right]\frac{\dot{\delta }(t)}{\delta (t)}+\frac{{{B}^{3}}}{2{{m}^{2}}\delta(t)}\int\limits_{0}^{t}{\frac{d{t}'}{{{\delta }^{2}}({t}')}}.
\end{eqnarray}

Upon integration, Equation (\ref{xdot}) yields {\it our main result}:
\begin{eqnarray} \label{eq:xfinal}
{{x}_{i}}(t)=X(t)+({{x}_{oi}}-{{X}_{o}})\frac{\delta (t)}{{{\delta }_{o}}}+\frac{{{B}^{3}}}{2{{m}^{2}}}\delta (t)\int\limits_{0}^{t}{\frac{d{t}'}{{{{\delta ^2}}}({t}')}}\int\limits_{0}^{{{t}'}}{\frac{d\tau }{{{\delta }^{2}}(\tau )}},
\end{eqnarray}
which constitutes the associated Bohmian trajectories of the Airy wave packet that satisfies Equation (\ref{eq:schroedinger1}). In particular, if $\delta (t) = 1$, $\omega(t) = 0$, 
Equation (\ref{eq:finalsol}) yields the result found by Berry and Balazs in Equation (\ref{eq:berry2}).\cite{berry} 

Furthermore, Equation (\ref{eq:xfinal}) displays still a myriad of nontrivial Airy packet trajectories. For example, if ${x_{oi}}$ is positive, then the particles distributed in the right half of the initial ensemble are accelerated whereas the particles distributed in the left half of the initial ensemble are decelerated. Besides, Equation (\ref{eq:xfinal}) implies that deviations from classical trajectories 
$\Delta {{x}_{i}}(t)={{x}_{i}}(t)-X(t)$ 
are entirely dependent on the solution of Equation (\ref{eq:deltaeq}) (which is a generalization of the Mathieu and Hill-Poschl-Teller equations).\cite{ruby}-\cite{nassar2}
Two independent solutions to Equation (\ref{eq:deltaeq}) can be obtained in general from just one particular solution to the same equation, namely,
\begin{eqnarray} \label{eq:delta1}
{{\delta }_{1}}(t)\equiv \delta (t)
\end{eqnarray}
and
\begin{eqnarray} \label{eq:delta2}
{{\delta }_{2}}(t)\equiv \delta (t)\int\limits_{{}}^{t}{\frac{dt'}{{{\delta }^{2}}(t')}}.
\end{eqnarray}

If initial conditions are imposed as follows: 
$\delta (0)=1$ and  $\dot{\delta }(0)=0$  
 we are led to the orthogonal conditions
${{\delta }_{1}}(0)=1$,  ${{\delta }_{2}}(0)=0$,  ${{\dot{\delta }}_{1}}(0)=0$ and  ${{\dot{\delta }}_{2}}(0)=1$. This constitutes a large set of general, nontrivial Bohmian trajectories associated to the Airy wave packet subject to time-dependent quadratic potentials. 

These new features are worth introducing to the subject's theoretical folklore in light of the fact that the evolution of a quantum mechanical Airy wave packet governed by the Schr\"odinger equation is analogous to the propagation of a finite energy Airy beam satisfying the paraxial equation. Numerous experimental configurations of optics and atom physics have shown that the dynamics of Airy beams depends significantly on initial parameters and configurations of the experimental set-up. The use of Airy beams for particle manipulation in nonlinear media remains a topic of intense theoretical and experimental research.\cite{siviloglou}-\cite{vallee}

 \pagebreak

{\bf Acknowledgments}

A. B. Nassar is very grateful to Prof. I. Besieris for his careful reading of the manuscript and numerous suggestions.
Part of this work was done at the UCLA
Physics and Astronomy Department. S. M-A grateful
acknowledges the MICINN (Spain) through Grant
FIS2011-29596-C02-01.


\begin{thebibliography}{99}
\bibitem{transform} These space-time transformations used here are a generalization of those found in references \cite{nassar,ray}. However, Equations (\ref{eq:xeq}) and (\ref{eq:deltaeq}) have clear physical meaning; they are essential in determining the trajectories of the Airy packet.
\bibitem{phase} The phase is given by: 

$\phi (x,t)=$

$=\frac{m{{x}^{2}}\dot{\delta }(t)}{2\delta }-\left[ \frac{mX(t)\dot{\delta }(t)}{\delta (t)}-m\dot{X}(t) \right]x+\frac{{{B}^{3}}}{2m}\left( \int\limits_{0}^{t}{\frac{dt'}{{{\delta }^{2}}(t')}} \right)\left( \frac{x-X(t)}{\delta (t)}-\frac{{{B}^{3}}}{6m}{{\left( \int\limits_{0}^{t}{\frac{dt'}{{{\delta }^{2}}(t')}} \right)}^{2}} \right)$

$-\int\limits_{0}^{t}{\left[ \frac{1}{2}m{{\delta }^{4}}(\tau ){{\left( \frac{X(\tau )\dot{\delta }(\tau )}{\delta {{(\tau )}^{2}}}-\frac{\dot{X}(\tau )}{\delta (\tau )} \right)}^{2}} \right]d\tau }$.
\end{thebibliography}
\end{document}